\documentstyle[11pt]{article}
\title{\bf COMPLEX $q$-ANALYSIS AND SCALAR FIELD THEORY
ON A $q$-LATTICE}
\author {\\ \\ \\ MARCELO R. UBRIACO\\
{\em Laboratory of Theoretical Physics}\\
{\em Department of Physics}\\
{\em P. O. Box 23343}\\
{\em University of Puerto Rico, R\'{\i}o Piedras}\\
{\em PR 00931-3343,  USA}}
\date{LTP-042-UPR\\February 1994}

\begin{document}
\vspace{0.5in}
\maketitle
\vspace{0.3in}
\begin{abstract}
We develop the basic formalism of complex $q$-analysis to study
the solutions of second order $q$-difference equations
which reduce, in the $q\rightarrow 1$ limit,
to the ordinary Laplace equation
in Euclidean and Minkowski space.  After
defining an inner product on the function space we construct
and study the properties of the solutions,
and then apply
this formalism to the Schr\"{o}dinger equation and
 two-dimensional scalar field theory.

\end{abstract}
\baselineskip20pt
\section{Introduction}
One of the most recently researched aspects of
 quantum group applications to physics has been
 on the role
 they could play as
symmetries in  quantum field theory .
For example, several  developments
on field theory with compact
quantum group internal symmetries can be
found in the literature \cite{AV,U2}.  Although
quantum versions of Lorentz and Poincar\'{e} algebras
 have been given by several
authors in refs. \cite{RT,PW}, the search for an
analytical formulation of  field theory on
quantum space-time  is still an open problem \cite{LNR}.
A starting point taken
by several authors  has been the study
of one-dimensional deformations of quantum mechanics
based on
$q$-deformations of the Heisenberg algebra. The analytical
framework for these approaches is based on
 $q$-analysis \cite{J} and the theory of
basic hypergeometric functions. In addition, as discussed in
references \cite{U2,LS},
$q$-analysis provides a concrete realization
of  $q$-deformed quantum mechanics as  standard
quantum mechanics on a non-uniform lattice.

In this paper
we extend the formalism given in Ref. \cite{U2} to two dimensions.
In particular, we consider the simplest case, which is
quantum mechanics and scalar field theory on a
complex $q$-plane with commuting coordinates.

Thus, the formalism
in this work gives  a preliminary insight
that could lead to more general formulations involving
quantum plane (non-commuting) coordinates. At the
beginning of  Sec. \ref{q-ana}
we introduce complex $q$-analysis by establishing a
correspondence between the $q$-plane derivatives and
difference operators, and  then we proceed by defining
an inner product consistent with the invariance of
the coordinate algebra under the adjoint operation.
 In Sec. \ref{q-or} we define a
second-order difference operator whose
 eigenfunctions are orthonormal with respect to
the inner product previously introduced. This result
allow us to expand a general solution in terms of
a $q$-analogue of the Fourier series.  In Sec \ref{QM}
 we apply this formalism to the Schr\"{o}dinger
equation, and in Sec. \ref{SCT} we construct the action
which leads to the  field equation
reducing, in the $q\rightarrow 1$ limit, to a scalar
field theory in two-dimensional Minkowski
space. In particular, the discreteness of
the spectrum is exhibited for the case $q\approx 0$.
\section{Complex $q$-analysis} \label{q-ana}
We start by considering two commuting coordinates
$\hat{x}^{i}$ satisfying the following relations with differentials
$d\hat{x}^{i}$ and derivatives $\hat{\partial}_{i}$
\begin{eqnarray}\label{CR}
\hat{x}^{i}d\hat{x}^{i}=p_{i}d\hat{x}^{i}\hat{x}^{i} \nonumber\\
\hat{\partial}_{i}\hat{x}^{i}=1+p_{i}\hat{x}^{i}\hat{\partial}_{i}\nonumber  \\
\hat{\partial}_{i}d\hat{x}^{i}=p_{i}^{-1}d\hat{x}^{i}\hat{\partial}_{i}
\nonumber\\
d\hat{x}^{i}d\hat{x}^{j}=-d\hat{x}^{j}d\hat{x}^{i},
\end{eqnarray}
where the two parameters $p_{i}$ are real numbers.
One can introduce a
$q$-complex number $\hat{x}\equiv \hat{x}^{1}$
and its adjoint $\overline{\hat{x}}\equiv \hat{x}^{2}$ if and only if
$p_{1}=p_{2}^{-1}\equiv q$.  One can check
that the relations in Eq. (\ref{CR}) can be written involving two matrices
$B$ and $C$ as follows
\begin{equation}
\hat{x}^{i}\hat{x}^{j}=B^{ij}_{kl}\hat{x}^{k}\hat{x}^{l} \nonumber
\end{equation}
\begin{equation}
\hat{\partial}_{j}\hat{x}^{i}=\delta^{i}_{j} +
C^{ik}_{jl}\hat{x}^{l}\hat{\partial}_{k}
\nonumber
\end{equation}
\begin{equation}
\hat{x}^{i}d\hat{x}^{j}=C^{ij}_{kl} d\hat{x}^{k}\hat{x}^{l} \nonumber
\end{equation}
\begin{equation}
\hat{\partial}_{j}d\hat{x}^{i}=(C^{-1})^{ik}_{jl}d\hat{x}^{l}\hat{\partial}_{k},
\nonumber
\end{equation}
where
\begin{equation}
B=\left(\begin{array}{cccc} 1 & 0 & 0 & 0 \\ 0 & 1-q & q & 0 \\
0 & 1 & 0 & 0 \\
0 & 0 & 0 & 1 \end{array}\right), \nonumber
\end{equation}
and
\begin{equation}
C=\left(\begin{array}{cccc} q & 0 & 0 & 0 \\ 0 & 0 & 1 & 0 \\
0 & 1 & 0 & 0 \\
0 & 0 & 0 & q^{-1} \end{array}\right) \nonumber
\end{equation}
are a particular choice of the matrices discussed in ref. \cite{BDR}.
The well known consistency conditions given in ref. \cite{WZ} are then
fulfilled
\begin{equation}
B_{12}B_{23}B_{12}=B_{23}B_{12}B_{23} \nonumber
\end{equation}
\begin{equation}
\left(B_{12}-{\bf 1}\right)\left(C_{12}+{\bf 1}\right)=0 \nonumber
\end{equation}
\begin{equation}
B_{12}C_{23}C_{12}=C_{23}C_{12}B_{23} \nonumber
\end{equation}
\begin{equation}
C_{12}C_{23}C_{12}=C_{23}C_{12}C_{23}.
\end{equation}
The invariance of the commutations relations in Equations (\ref{CR}) is
restricted to the group
of scale transformations
\begin{equation}
t=\left(\begin{array}{cc} a & 0 \\ 0 & d \end{array}\right) ,
\end{equation}
with $[a,d]=0$. Since the coordinate $\hat{x}$
and its adjoint commute, we can now
identify  $\hat{x}$
with the usual complex coordinate $w$, such that to the adjoint
$\overline{\hat{x}}$ will correspond the complex conjugate $x^{\ast}$.
Thus, the $q$-derivatives $\hat{\partial}\equiv\hat{\partial}_{1}$
{}~, $\overline{\hat{\partial}}\equiv\hat{\partial}_{2}$ and differentials
$d\hat{x}$,$d\overline{\hat{x}}$ are realized on the usual complex plane in
terms of  $q$-difference operators according to the following
relations
\begin{equation}
\hat{\partial}:D_{+}=w^{-1}\frac{1-T}{1-q}
\end{equation}
\begin{equation}
\overline{\hat{\partial}}:D_{-}^{\ast}=w^{\ast-1}\frac{1-T^{\ast-1}}
{1-q^{-1}}
\end{equation}
\begin{equation}
d\hat{x}:d_{q}w\, T^{-1}
\end{equation}
\begin{equation}
d\overline{\hat{x}}:d_{q}w^{\ast}\, T^{\ast},
\end{equation}
where $T=q^{w\partial_{w}}$ is the scaling operator and hereafter
$0<q<1$.  Equations (\ref{CR})
written in terms of the variables $w$ and $w^{\ast}$ are  consistent
if we can define an inner product such the adjoint of the $q$-difference
operators and $q$-differentials satisfy
\begin{equation}
\overline{D_{+}}=-q^{-1}D_{-}^{\ast} \label{q-her}
\end{equation}
\begin{equation}
\overline{d_{q}w}\propto d_{q}w^{\ast}.
\end{equation}
In fact, Equation (\ref{q-her}) is satisfied if we define our inner product
in terms of a double $q$-integral
\begin{equation}
<\phi,\psi>\equiv \int_{}^{}d_{q}w  d_{q}w^{\ast} \phi^{\ast}\psi ,
\label{innpro}
\end{equation}
provided that the space functions satisfy the appropriate
boundary conditions such that the boundary term
\begin{equation}
\int_{}^{} d_{q}w
d_{q}w^{\ast}D_{+}^{\ast}\left(\phi^{\ast}T^{\ast-1}\psi\right)=0.
\label{bt}
\end{equation}
 This inner product corresponds to
the simplest two-dimensional generalization  of the one defined in \cite{U1}.
By introducing an independent complex variable $z$  on a  linear lattice
and considering the coordinate $w$ as a function $w(z)=q^{z}$,
Eq. (\ref{innpro}) can be interpreted as an equation taking values on
a non-uniform lattice ($q$-lattice) of step size
$\Delta w(z)\equiv w(z)-w(z+1)=(1-q)w(z)$.  Explicitly, Eq. (\ref{innpro})
reads
\begin{equation}
<\phi,\psi>=\sum_{n=0}^{\infty}\sum_{m=0}^{\infty}\Delta w_{n}
\Delta w_{m}^{\ast} \phi^{\ast}(w_{n},w_{m}^{\ast})\psi(w_{n},w_{m}^{\ast}),
\end{equation}
where $w_{n}\equiv q^{n}w$ is the point after either $n$ scalings in the
$q$-lattice or $n$  unit translations in the linear lattice.
In fact, for $q=e^{-\lambda}$ very close to one we
can approximate at first order
in $\lambda$ as
$q\approx 1-\lambda$ and the $q$-difference operators become proportional
to difference operators on a linear lattice.
\section{Orthogonal functions} \label{q-or}
Starting with the $q$-difference operators discussed in the previous
section, there is no unique second order $q$-difference operator
one could write which would lead to the
ordinary laplacian as $q\rightarrow 1$.  Some aspects of this
lack of uniqueness have been discussed in \cite{U1} wherein
a hermitian free hamiltonian was found to be given by the operator
\begin{equation}\label{h}
h_{0}=-x^{-1}[x\partial]x^{-1}[x\partial_{x}],
\end{equation}
with $[x\partial]\equiv \frac{T^{1/2}-T^{-1/2}}{q^{1/2}-
q^{-1/2}}$ \label{op1}.
Eigenfunctions of this operator are the basic exponentials
$E(\sqrt{q};x)$ defined by the formula \cite{Ex}
\begin{equation}
E(\sqrt{q};x)=\sum_{n=0}^{\infty}\frac{x^{n}}{[n]!}.
\end{equation}
But in contrast to the (continuous) $q=1$ case and the one
(linear lattice) with exponentials with integer momentum,
the basic exponentials are not orthogonal with respect to
an inner product  of the type given in Eq.(\ref{innpro}).
In fact, the basic sine $S(x)$ and cosine $C(x)$ functions
defined by
\begin{equation}
E(\sqrt{q},ix)=C(x)+i S(x), \label{e}
\end{equation}
satisfy in the interval $[-x\prime,x\prime]$ the following
orthogonality relations
\begin{equation}
\int_{-x\prime}^{x\prime} d_{q}x  S(\kappa x) S(\kappa' x)=
M(\kappa,x\prime)\delta_{\kappa,\kappa'} \label{ss}
\end{equation}
\begin{equation}
\int_{-x\prime}^{x\prime} d_{q}x  S(\kappa x) C(\kappa'x)=0
\label{sc}
\end{equation}
\begin{equation}
\int_{-x\prime}^{x\prime} d_{q}x  C(\sqrt{q}\kappa x) C(\sqrt{q}\kappa'x)=
N(\sqrt{q}\kappa,x\prime)\delta_{\kappa,\kappa'} , \label{cc}
\end{equation}
where  $M$ and $N$ denote normalizations and the
 allowed values of $\kappa$ are those
which satisfy $S(\kappa x\prime)=0$.
  The generalization of the operator in Eq. (\ref{h})
to the two-dimensional euclidean case should lead
in the $q\rightarrow 1$ limit
to the laplacian $\Delta=\partial_{w}\partial_{w^{\ast}}$.
Although this generalization is in principle non
unique, we see that
according to Eqs.(\ref{ss}),(\ref{sc}) and (\ref{cc}) its
eigenfunctions should correspond to the $q$-analogue of either
$sin(\kappa^{\ast}w+\kappa w^{\ast})$ or
$cos(\kappa^{\ast}w+\kappa w^{\ast})$. In order to
obtain a $q$-analogue of
these functions we define the symbol
\begin{equation}
(a,b)^{(n)}\equiv\sum_{m=0}^{n}\frac{[n]!}{[m]![n-m]!}
a^{m}b^{n-m}=(b,a)^{(n)} ,
\end{equation}
which can be check to satisfy the following difference equation
\begin{equation}
a^{-1}[a\partial_{a}](a,b)^{(n)}=[n](a,b)^{(n-1)}.
\end{equation}
This symbol vanishes for $b=-aq^{\pm 1/2}$ with $n$ even, and
$b=-a$ with $n$ odd. In particular
\begin{equation}
(a,b)^{(1)}=a+b, \nonumber
\end{equation}
and
\begin{eqnarray}
(a,b)^{(2)}&=&a^{2}+[2]ab+b^{2} \\ \nonumber
&=&(aq^{-1/2},b)^{(1)}(aq^{1/2},b)^{(1)} \nonumber
\end{eqnarray}
is a $q$-analogue of the binomial theorem.
Therefore, the function
\begin{equation}
S(\kappa w^{\ast},\kappa^{\ast}w)\equiv
\sum_{n=0}^{\infty}\frac{(-1)^{n}(\kappa w^{\ast},\kappa^{\ast}w)
^{(2n+1)}}{[2n+1]!},
\end{equation}
is an eigenfunction of the operator
\begin{equation} \label{nabla}
\nabla_{q}^{2}\equiv
w^{-1}[w\partial_{w}]w^{\ast-1}[w^{\ast}\partial_{w^{\ast}}] ,
\label{op2}
\end{equation}
satisfying the difference equation
\begin{equation}
-\nabla_{q}^{2}S(\kappa w^{\ast},\kappa^{\ast}w)=
\kappa\kappa^{\ast}S(\kappa w^{\ast},\kappa^{\ast}w).
\end{equation}
The orthogonality of these functions follow from the simple
identity
\begin{equation}
S(a,b)=S(a) C(b)
+ C(a) S(b), \label{s}
\end{equation}
and the orthogonality relations in Eqs. (\ref{ss}), (\ref{sc})
and (\ref{cc}). This function can be generalized to include more
variables. For example, introducing a new symbol
\begin{equation}
\left((a,b),(c,d)\right)^{(n)}=\sum_{m=0}^{n}
\frac{[n]!}{[m]![n-m]!}(a,b)^{(m)}(c,d)^{(n-m)},
\end{equation}
one can define the function
\begin{equation}
S\left((a,b),(c,d)\right)\equiv\sum_{n=0}\frac{(-1)^{n}
\left((a,b),(c,d)\right)^{(2n+1)}}{[2n+1]!},
\end{equation}
such that
\begin{equation}
S\left((a,b),(c,d)\right)=S(a,b) C(c,d)+C(a,b) S(c,d).
\end{equation}
In the two-dimensional case, restricting the values of $\kappa$ to those
which satisfy $S(\kappa^{\ast}w\prime)=0$ we write
\begin{equation}
\int_{-w\prime}^{w\prime}\int_{-w\prime^{\ast}}^{w\prime^{\ast}}
d_{q}w^{\ast}d_{q}w\,S(\kappa w^{\ast},\kappa^{\ast}w)
 S(\kappa' w^{\ast},\kappa'^{\ast}w)
= 2 M(\kappa,w\prime^{\ast}) N(\kappa^{\ast},w\prime)\delta_{\kappa,\kappa'},
\end{equation}
where the values of $M$ and $N$ have to be computed numerically
for a particular choice of $q$. Since the zeros of the
function $S(x)$ are real,
$\kappa w\prime^{\ast}= \kappa^{\ast} w\prime$, these normalizations satisfy
$M(\kappa,w\prime^{\ast}) N(\kappa^{\ast},w\prime)=
M(\kappa^{\ast},w\prime) N(\kappa,w\prime^{\ast})$.  On the other hand,
the identity
\begin{equation}
C(a,b)=C(a)C(b)-S(a)S(b) ,\label{c}
\end{equation}
shows that if $\kappa w\prime^{\ast}$ is a zero of either $S(x)$
or $C(x)$, the functions $C(\kappa w^{\ast},\kappa^{\ast}w)$
are not orthogonal.
Thus, the only $q$-orthonormal eigenfunctions
of the operator in Eq.(\ref{op2}) are given by the set
\begin{equation} \label{f}
f_{\kappa}(w,w^{\ast})=\frac{1}{\sqrt{2
M(\kappa,w\prime^{\ast}) N(\kappa^{\ast},w\prime)}}
S(\kappa w^{\ast},\kappa^{\ast}w).
\end{equation}
{}From these properties, we can formally expand an
arbitrary function \\ $F(w,w^{\ast})$ as a $q$-analogue of a
Fourier series
\begin{equation}
F(w,w^{\ast})=\sum_{\kappa>0}^{}A(\kappa)f_{\kappa}(w,w^{\ast}),
\end{equation}
where the coefficients $A(\kappa)$ are given
by the formula
\begin{equation}\label{A}
A(\kappa)=\int_{-w\prime}^{w\prime}\int_{-w\prime^{\ast}}^{w\prime^{\ast}}
d_{q}w d_{q}w^{\ast} F(w,w^{\ast})f_{\kappa}(w,w^{\ast}).
\end{equation}
\section{Quantum Mechanics}\label{QM}
In this section, based on the discussion in Section (\ref{q-or})
, we solve the Schr\"{o}dinger equation
 and show the closure property of its solutions.
Introducing a standard time variable, the
time dependent  Schr\"{o}dinger equation
becomes the following difference equation
\begin{equation}
-\nabla_{q}^{2}\Psi(w,w^{\ast},t)=i\partial_{t}\Psi(w,w^{\ast},t).
\end{equation}
The corresponding stationary state solutions
\begin{equation}
f_{\kappa}(w,w^{\ast})e^{-iE_{\kappa}t}
\end{equation}
are written
in terms of the $q$-orthonormal functions given in Equation
(\ref{f}). As already discussed in the previous section,
the set of $\kappa$ values are those which make the
functions $f_{\kappa}$ to vanish at the boundary points.
Then, a general solution of the time dependent equation can be
expanded as
\begin{equation}\label{exp}
\Psi(w,w^{\ast},t)=\sum_{\kappa>0}^{}c_{\kappa}
f_{\kappa}(w,w^{\ast})e^{-iE_{\kappa}t},
\end{equation}
with eigenvalues $E_{\kappa}=\kappa\kappa^{\ast}$.
According to Equation (\ref{A}) the coefficients are given
by the inverse transformation
\begin{eqnarray}
c_{\kappa}&=&2\sum_{n,m=0}^{\infty}\Delta w\prime_{n}\Delta w\prime_{m}^{\ast}
\Psi(w\prime_{n},w\prime_{m}^{\ast},0)f_{\kappa}(w\prime_{n},w\prime_{m}^{\ast}) \nonumber \\
& & +2\sum_{n,m=0}^{\infty}\Delta w\prime_{n}\Delta w\prime_{m}^{\ast}
\Psi(w\prime_{n},-w\prime_{m}^{\ast},0)f_{\kappa}(w\prime_{n},-w\prime_{m}^{\ast}),
\end{eqnarray}
such that replacing back into Equation (\ref{exp}), a simple
algebraic manipulation leads to the closure relation for the
functions $f_{\kappa}$
\begin{equation}
\sum_{\kappa.0}f_{\kappa}(w\prime_{n},w\prime_{m}^{\ast})
f_{\kappa}(w\prime_{r},w\prime_{s}^{\ast})=\frac{\delta_{n,r}\delta_{m,s}}{2\Delta w\prime_{n}
\Delta w\prime_{m}}.
\end{equation}
The kernel $K(w\prime_{n},w\prime_{m}^{\ast};w\prime_{r},w\prime_{s}^{\ast};t)$
is therefore given by
\begin{equation}
K(w\prime_{n},w\prime_{m}^{\ast};w\prime_{r},w\prime_{s}^{\ast};t)=
\sum_{\kappa>0}f_{\kappa}(w\prime_{n},w\prime_{m}^{\ast})
f_{\kappa}(w\prime_{r},w\prime_{s}^{\ast})e^{iE_{\kappa}t}
\end{equation}
and relates solutions of the free equation according to
\begin{equation}
\Psi(w\prime_{r},w\prime_{s}^{\ast};t)=\sum_{n,m=0}\Delta w\prime_{n}
\Delta
w\prime_{m}^{\ast}K(w\prime_{n},w\prime_{m}^{\ast};w\prime_{r},w\prime_{s}^{\ast};t)
\Psi(w\prime_{n},w\prime_{m}^{\ast};0).
\end{equation}
\section{Scalar Field Theory}\label{SCT}
In this section we apply the formalism discussed in sections
(\ref{q-ana}) and (\ref{q-or}) to study some basic
aspects of scalar field theory on a $q$-lattice. In particular,
we are interested in the case which corresponds in the
$q\rightarrow 1$ limit to a
field theory in two-dimensional Minkowski space.  In order to
resolve the problem of uniqueness, it is natural
to require that the field equation  be derived from an action, in
a similar way as it is done for the $q=1$ case.  Therefore, based
on the discussion in Section (\ref{q-ana}), we define
a exterior derivative $d$ in terms of the right derivatives
$D_{+}$ and $D_{+}^{\ast}$ as follows
\begin{equation}
d=d_{q}w D_{+}+d_{q}w^{\ast}D_{+}^{\ast} .
\end{equation}
The lagrangian density ${\cal L}$ for a real scalar field
can therefore be defined in terms of the
usual Hodge operation $\ast d_{q}w=id_{q}w^{\ast}$ such that
\begin{equation}
d_{q}wd_{q}w^{\ast}{\cal L}\equiv -iq^{1/2}d\phi\ast d\phi ,
\end{equation}
and the action  becomes
\begin{eqnarray}
S&=&\frac{q^{1/2}}{2}\int_{}^{}\int_{}^{} d_{q}w d_{q}w^{\ast}
\left[[D_{+}\phi(w,w^{\ast})]^{2}+
[D_{+}^{\ast}\phi(w,w^{\ast})]^{2}\right]\nonumber \\
&=&\!\!\!\!\frac{q^{1/2}}{2}\sum_{n,m=0}^{\infty} \Delta w_{n}\Delta
w_{m}^{\ast}
\left[\left(\frac{\Delta\phi(w_{n},w_{m}^{\ast})}{\Delta w_{n}}\right)
^{2}\!\!+\!\left(\frac{\Delta^{\ast}\phi(w_{n},w_{m}^{\ast})}
{\Delta w_{n}^{\ast}}\right)^{2}\right],
\end{eqnarray}
where it is understood that $S$ is being valued at the
boundaries.  Variation of the action with respect to $\phi$ results in the
field equation
\begin{equation}
 - \left[\left(w^{-1}[w\partial_{w}]\right)^{2}+c.c.\right]\phi=0, \label{feq}
\end{equation}
corresponding in the $q\rightarrow 1$ limit
to the flat space equation
$\left(-\partial_{x}^{2}+\partial_{y}^{2}\right)\phi=~0$. A solution of the
field equation
(\ref{feq}) can be written as a sum in terms of the orthonormal sets
\begin{equation}
g_{\kappa}(w,w^{\ast})=\frac{1}{{\sqrt{\cal M}}}
S(\kappa w,\kappa^{\ast}w^{\ast}),
\end{equation}
and
\begin{equation}
h_{\kappa}(w,w^{\ast})=\frac{1}{{\sqrt{\cal N}}}
\tilde{C}(\kappa w,\kappa^{\ast}w^{\ast})
\end{equation}
where the normalizations are written in terms of the values $M$ and $N$,
as displayed in Eqs. (\ref{ss}) and (\ref{cc}). The function
$\tilde{C}(a,b)$
is obtained by conveniently modifying the function $C(a,b)$ in
Eq.(\ref{c}) such that
\begin{equation}
\tilde{C}(a,b)=C(\sqrt{q}a)C(\sqrt{q}b)-S(a)S(b). \label{ctilde}
\end{equation}
Therefore, we define the real field $\phi$
\begin{equation}
\phi(w,w^{\ast})=\sum_{\kappa\geq 0}\left[A_{\kappa}g_{\kappa}(w,w^{\ast})
+B_{\kappa}h_{\kappa}(w,w^{\ast})\right],\label{phi}
\end{equation}
where the values of $\kappa$ satisfy $S(\kappa w\prime)=0$.  The real
coefficients $A_{\kappa}$ and $B_{\kappa}$ can be written in terms
of the inverse transformations
\begin{equation}
A_{\kappa'}=\int_{-w\prime}^{w\prime}\int_{-w\prime^{\ast}}^{w\prime^{\ast}}
d_{q}w d_{q}w^{\ast}g_{\kappa}(w,w^{\ast}) g_{\kappa'}(w,w^{\ast}),
\end{equation}
and
\begin{equation}
B_{\kappa'}=\int_{-w\prime}^{w\prime}\int_{-w\prime^{\ast}}^{w\prime^{\ast}}
d_{q}w d_{q}w^{\ast} h_{\kappa}(w,w^{\ast}) h_{\kappa'}(w,w^{\ast}).
\end{equation}
Without loss of generality  we can define
\begin{equation}
\kappa=\frac{k_{0}-ik_{1}}{\sqrt{2}}\equiv\frac{\alpha}{w\prime} ,
\end{equation}
where $\alpha\in{\bf R}$ are zeros of $S(x)$.  The
condition $\kappa^{2}+\kappa^{\ast 2}=0$  relates
the boundary points such that $w\prime^{\ast}=iw\prime$.
As a concrete example, we consider the case $q\approx 0$.
In this case, a very good approximation to the zeros of $S(x)$ and $C(x)$
is given \cite{Ex} respectively  by the set of values
\begin{equation}
\alpha_{m}=\frac{q^{-m+1/4}}{1-q} ,
\end{equation}
and
\begin{equation}
\beta_{m}=\frac{q^{-m+3/4}}{1-q} ,
\end{equation}
where $m=1,2,3,...$.  From the definition of the
function $\tilde{C}$ in Eq. (\ref{ctilde}) and the relation
between the zeros $\beta_{m}=\sqrt{q}\alpha_{m}$, we see that
the expansion
for the field at the lattice point $(w_{n}\prime,w_{m}\prime^{\ast})$,
$n\leq m$, reduces to the finite sum
\begin{equation}
\phi(w_{n}\prime,w_{m}\prime^{\ast})=\sum_{l=1}^{m}\left[
A_{l} S\left(\frac{q^{-l+1/4}(q^{m},q^{n})}{1-q}\right) +
B_{l} \tilde{C}\left(\frac{q^{-l+1/4}(q^{m},q^{n})}{1-q}\right)\right],
\end{equation}
and therefore $\phi(w\prime,w\prime^{\ast})=0$.
The corresponding discrete spectrum $k_{0}$ is given by
\begin{equation}
k_{0}=\pm \frac{q^{-l+1/4}}{\sqrt{2|w\prime|}(1-q)}.
\end{equation}
\section{Discussion}
In this letter, we have generalized the formalism
of real $q$-analysis to the complex case. We constructed
the solutions of second order $q$-difference operators
which in the $q\rightarrow 1$ limit reduce to
the laplacians in Euclidean and Minkowski space.  In order
to understand the meaning of the
adjoint operation and therefore to study the basic properties
 of these solutions we
defined, generalizing a previous work \cite{U1}, an
inner product as a double $q$-integral.  This inner product
introduces a $q$-lattice structure and it
specifies the type of solutions that satisfy the
orthogonality and closure properties. We have seen that
given the solutions for any of the
two $q$-difference operators here considered, most
of these solutions will fail to form an orthogonal set.
This case contrasts with the case of a linear
lattice, $q\approx 1$, wherein the sine, cosine and
exponential functions are all invariant under lattice
translations,
therefore sharing the orthogonality and closure properties.
We illustrated our results by considering
 the Schr\"{o}dinger equation for the Euclidean case and
scalar field theory for the Minkowski case
 . From the point of
view of noncommutative geometry, this work provides a
preliminary basis
to search for an analytical approach to
 quantum mechanics and field theory involving
quantum plane coordinates.
On the other hand, by taking several copies of the
coordinate pair $(w,w^{\ast})$, one could generalize
this work to a higher dimensional $q$-lattice with
the goal that it could be used as a a tool to regularize standard
field theory.  It would be of much interest to check if
some of the problems that arise when one
defines a Dirac action on a linear lattice still
remain when a more general geometry such as a $q$-lattice is considered.

\end{document}